\documentclass[12pt]{article}
\begin{document}
\baselineskip=16pt

\title
{\begin{Huge}
The Horizon Energy of a Black Hole\\
\end{Huge}
\vspace{0.5in} }
\author{Yuan K. Ha\\ Department of Physics, Temple University\\
 Philadelphia, Pennsylvania 19122 U.S.A. \\
 yuanha@temple.edu \\    \vspace{.1in}  }
\date{December 1, 2017}
\maketitle
\vspace{.3in}
\begin{abstract}
\noindent
We investigate the energy distribution of a black hole in various spacetimes as reckoned by a distant
observer using the quasi-local energy approach. In each case the horizon mass of a black hole: 
neutral, charged or rotating, is found to be twice the irreducible mass observed at infinity. This is
known as the Horizon Mass Theorem. As a consequence, the electrostatic energy and the rotational 
energy of a general black hole are all external quantities. Matter carrying charges and spins could
only lie outside the horizon. This result could resolve several long-standing paradoxes related to
known black hole properties; such as why entropy is proportional to area and not to volume, the 
information loss problem, the firewall problem, the internal structure and the thin shell model of a
black hole.\\

\end{abstract}

\newpage
\vspace*{.5in}
\noindent
{\bf 1. \hspace{0in} Quasi-local Energy}\\

\vspace{.2in}
\noindent
A black hole has the strongest gravitational field of all gravitational systems and the
greatest gravitational potential energy. It is well known that gravitational energy density
cannot be defined consistently in general relativity since gravitational field can be
transformed away in a local inertial frame. Nevertheless, it is possible to consider the
total energy contained in a surface enclosing a black hole at a given coordinate distance.
This is based on the quasi-local energy approach [1] obtained from a Hamiltonian-Jacobi
analysis of the Hilbert action in general relativity.\\
\indent
   The quasi-local energy expression is the most important development in general
relativity in recent years to understand the dynamics of the gravitational field, such as
energy, momentum and angular momentum [2]. For asymptotically flat spacetime, the
quasi-local energy agrees with the Arnowitt-Deser-Misner energy [3] at spatial infinity
and for spherically symmetric spacetime, it has the correct Newtonian limit, including
negative contribution to gravitational binding. It also agrees with the Komar energy [4]
and Bondi energy [5] at null infinity. The quasi-local energy approach is therefore 
naturally suited for investigating the energy distribution of a black hole.\\
\indent
   The expression for the quasi-local energy is given in terms of the total mean curvature
of a surface bounding a volume for a gravitational system in four-dimensional spacetime.
The Brown and York expression is given in the form of an integral [1]
\begin{equation}
E = \frac{c^{4}}{8\pi G} \int_{^{2}B} d^{2}x \sqrt{\sigma} (k-k^{0}) ,
\end{equation}
where $\sigma$ is the determinant of the metric defined on the two-dimensional surface $^{2}B$ ;
$k$ is the trace of extrinsic curvature of the surface and $k^{0}$ , the trace of curvature of a
reference space. For asymptotically flat reference spacetime, $k^{0}$ is zero.\\

\newpage
\noindent
{\bf 2. \hspace{0in} Horizon Mass Theorem}\\

\noindent
The Horizon Mass Theorem is the final outcome of the quasi-local energy expression
applied to the black hole. The mass of a black hole depends on where the observer is. The
closer one gets to a black hole the less gravitational energy one expects to see. As a
result, the mass of a black hole increases as one gets near the horizon. The Horizon Mass
Theorem can be stated in the following [6]:\\

\noindent
{\bf Theorem.} For all black holes: neutral, charge or rotating, the horizon mass is always
twice the irreducible mass observed at infinity.\\

\noindent
It is useful to introduce the following definitions of mass in order to understand the
energy of a black hole:\\

\begin{enumerate}
  \item The {\em asymptotic mass} is the mass of a neutral, charged or rotating black hole
        including electrostatic and rotatinal energy. It is the mass observed at infinity.
  \item The {\em horizon mass} is the mass which cannot escape from the horizon of a neutral,
        charged or rotating black hole. It is the mass observed at the horizon.
  \item The {\em irreducible mass} is the final mass of a charged or rotating black hole when
        its charge or angular momentum is removed by adding external particles to the 
        black hole. It is the mass observed at infinity.
\end{enumerate}

\vspace{.2in}
\noindent
{\bf 3. \hspace{0in} Schwarzschild Black Hole}\\

\noindent
The total energy contained in a sphere enclosing the black hole at a coordinate distance $r$
is given by the expression [1,7,8]\\
\begin{equation}
E(r) = \frac{rc^{4}}{G} \left[ 1- \sqrt{ 1- \frac{2GM}{rc^{2}} } \right],
\end{equation}
where $M$ is the mass of the black hole observed at infinity, $c$ is the speed of light and $G$ is
the gravitational constant. At the horizon, the Schwarzschild radius is $r = 2GM/c^{2}$ .
Evaluating the expression in Eq.(2), we find the metric coefficient 
$g_{00}=(1-2GM/rc^{2})^{1/2}$ vanishes identically and the energy at the horizon is therefore
\begin{equation}
E(r)=\left( \frac{2GM}{c^{2}} \right) \frac{c^{4}}{G} = 2Mc^{2} .
\end{equation}
The horizon mass of the Schwarzschild black hole is simply twice the asymptotic mass
$M$ observed at infinity. The negative gravitational energy outside the black hole is as
great as the asymptotic mass.\\

\vspace{.2in}
\noindent
{\bf 4. \hspace{0in} Charged Black Hole}\\

\noindent
The total energy of a charged black hole contained within a radius at coordinate $r$ is now
given by [7]
\begin{equation}
E(r) = \frac{rc^{4}}{G} \left [ 1 - \sqrt{ 1 - \frac{2GM}{rc^{2}} + \frac{G Q^{2}}{r^{2}c^{4}} } \right ] ,
\end{equation}
where $M$ is the mass of the black hole including electrostatic energy observed at infinity
and $Q$ is the electric charge. At the horizon radius
\begin{equation}
r_{+} = \frac{GM}{c^{2}} + \frac{GM}{c^{2}} \sqrt{1 - \frac{Q^{2}}{GM^{2}} } ,
\end{equation}
the square root in Eq.(4) again vanishes and the horizon energy becomes
\begin{equation}
E(r_{+}) = \frac{r_{+}c^{4}}{G} = Mc^{2} + Mc^{2} \sqrt{ 1 - \frac{Q^{2}}{GM^{2}} }.
\end{equation}
When this is expressed in terms of the irreducible mass of the charged black hole
\begin{equation}
M_{irr} = \frac{M}{2} + \frac{M}{2} \sqrt{ 1 - \frac{Q^{2}}{GM^{2}} },
\end{equation}
the horizon energy becomes exactly twice the irreducible energy
\begin{equation}
E(r_{+}) = 2M_{irr}c^{2} .
\end{equation}
The horizon mass therefore depends only on the energy of the black hole when it is
neutralized by adding oppositely charged particles. There is no electrostatic energy inside
the charged black hole.\\

\vspace{.2in}
\noindent
{\bf 5. \hspace{0in} Slowly Rotating Black Hole}\\

\noindent
The total energy of a slowly rotating black hole with angular momentum $J$ and angular
momentum parameter $a = J/Mc$ using the quasi-local energy approach is given by the
approximate expression [9], $0 < a \ll 1$ ,
\begin{eqnarray}
E(r) & = & \frac{rc^{4}}{G} \left[ 1 - \sqrt{ 1 - \frac{2GM}{rc^{2}} + \frac{a^{2}}{r^{2}} } \right ] \nonumber \\
     &   & + \frac{a^{2}c^{4}}{6rG} \left [ 2 + \frac{2GM}{rc^{2}} 
        + \left ( 1 + \frac{2GM}{rc^{2}} \right ) \sqrt{ 1 - \frac{2GM}{rc^{2}} + \frac{a^{2}}{r^{2}} } \right ] + \cdots    \end{eqnarray}
Again, using the horizon radius in this case
\begin{equation}
r_{h} = \frac{GM}{c^{2}} + \sqrt { \frac{G^{2}M^{2}}{c^{4}} - \frac{J^{2}}{M^{2}c^{2}} }
\end{equation}
and the irreducible mass
\begin{equation}
M_{irr}^{2}  = \frac{M^{2}}{2} + \frac{M^{2}}{2} \sqrt { 1 - \frac{J^{2}c^{2}}{G^{2}M^{4}} } ,
\end{equation}
we arrive at a very good approximate relation for the horizon energy
\begin{equation}
M_{h} \simeq 2 M_{irr} + O(a^{2}) . 
\end{equation}

\vspace{.2in}
\noindent
{\bf 6. \hspace{0in} Black Hole at Any Rotation}\\

\noindent
For general and fast rotations, the quasi-local energy approach has limitation but the total
energy can be obtained very accurately by numerical evaluation in the teleparallel 
formulation of general relativity [10]. The teleparallel gravity is an equivalent geometric
formulation of general relativity in which the action is constructed purely with torsion
without curvature. It has a gauge field approach. There is a perfectly well-defined
gravitational energy density and the result agrees very well with Eq.(12) at any rotation.
The small discrepancy is due to the axial symmetry of a rotating black hole compared
with the exact spherical symmetry of a Schwarzschild black hole. The result shows that
the rotational energy appears to reside almost completely outside the black hole.\\
\indent
   For an exact relationship, however, we have to employ a formula known for the area of
a rotating black hole valid for all rotations in the Kerr metric [11],
\begin{equation}
A = 4 \pi ( r_{h}^{2} + a^{2} ) = \frac {16 \pi G^{2}M_{irr}^{2}}{c^{4}} .
\end{equation}
This area is exactly the same as that of a Schwarzschild black hole with asymptotic mass
$M_{irr}$ . Now a local observer who is comoving with the rotating black hole at the event
horizon will see only this Schwarzschild black hole. Since the horizon mass of the
Schwarzschild black hole is $2M_{irr}$ , therefore the horizon mass of the rotating black hole
is exactly $M_{h} = 2M_{irr}$ . We have shown in each case, the horizon mass of a black hole
is always twice the irreducible mass observed at infinity.\\
\indent
   The Horizon Mass Theorem shows that the electrostatic energy and the rotational 
energy of a general black hole are all external quantities. They are absent inside the black
hole. A charged black hole does not have electric charges inside. A rotating black hole
does not rotate; only the external space is rotating. The conclusion is surprising. It could
resolve several long-standing paradoxes of black holes such as the entropy problem, the
information problem, the firewall problem and the gravastar thin shell model since matter
carrying charges and spins could only stay outside the horizon. The quasi-local energy of
black holes is one of the profound and fascinating results known recently in general
relativity.\\


\newpage
\begin{thebibliography}{99}
\bibitem{1} J.D. Brown and J.W. York, Jr. {\em Phys. Rev. D} {\bf 47}, 1407 (1993).     
\bibitem{2} M.T. Wang and S.T. Yau, {\em Phys. Rev. Lett.} {\bf 102}, 021101 (2009).    
\bibitem{3} R. Arnowitt, S. Deser and C.W. Misner, {\em Phys. Rev.} {\bf 117}, 1595 (1960).
\bibitem{4} A. Komar, {\em Phys. Rev.} {\bf 113}, 934 (1959).           
\bibitem{5} H. Bondi, M.G.J. van der Burg and A.W.K. Metzner, {\em Proc. R. Soc. London Ser. A} {\bf 269}, 21 (1962).
\bibitem{6} Y.K. Ha, {\em Int. J. Mod. Phys.} D {\bf 14}, 2219 (2005).        
\bibitem{7} J.W. Maluf, {\em J. Math. Phys.} {\bf 36}, 4242 (1995).             
\bibitem{8} Y.K. Ha, Gen. Rel. Gra. {\bf 35}, 2045 (2003).        
\bibitem{9} E.A. Martinez, {\em Phys. Rev.} D{\bf 50}, 4920 (1994).            
\bibitem{10} J.W. Maluf, E.F. Martins and A. Kneip, {\em J. Math. Phys.} {\bf 37}, 6302 (1996).                      
\bibitem{11} D. Christodoulou and R. Ruffini, {\em Phys. Rev. D} {\bf 4}, 3552 (1971).       
\end{thebibliography}
\end{document}